\documentclass[twocolumn]{aastex631}

\usepackage{amsmath,amssymb,amstext}
\usepackage{tablefootnote}

\begin{document}
\title{Numerical  models of planetary nebulae with different episodes of mass ejection: the particular case of HuBi 1.}
\author{Ary Rodr\'iguez-Gonz\'alez}
\affiliation{Instituto de Ciencias Nucleares, Universidad Nacional Aut\'onoma de M\'exico\\
Ap. 70-543, 04510, Ciudad de M\'exico, M\'exico}
\author{Miriam Pe\~na}
\affiliation{Instituto de Astronom\'ia, Universidad
Nacional Aut\'onoma de M\'exico\\
Ap. 70-264, 04510, Ciudad de M\'exico, M\'exico}

\author{Liliana Hern\'andez-Mart{\'\i}nez}
\affiliation{Facultad de Ciencias.  Universidad Nacional Aut\'onoma de M\'exico\\
Ap. 70-399, 04510 Ciudad de M\'exico, M\'exico}
\affiliation{Escuela Superior de Física y Matemáticas, Instituto Politécnico Nacional, U.P. Adolfo López Mateos, C.P. 07738,
Ciudad de México, México}

\author{Francisco Ruiz-Escobedo}
\affiliation{Instituto de Astronom\'ia, Universidad
Nacional Aut\'onoma de M\'exico\\
Ap. 70-264, 04510, Ciudad de M\'exico, M\'exico}

\author{Alejandro Raga\dag}
\affiliation{Instituto de Ciencias Nucleares, Universidad Nacional Aut\'onoma de M\'exico\\
Ap. 70-543, 04510, Ciudad de M\'exico, M\'exico}
\author{Grazyna Stasi\'nska}
\affiliation{Laboratoire Univers et Th\'eorie, Observatoire de Paris, Universit\'e PSL\\
Universit\'e
Paris Cit\'e, CNRS, F-92190 Meudon, France}
\author{Jorge Ivan Castorena}
\affiliation{Instituto de Ciencias Nucleares, Universidad Nacional Aut\'onoma de M\'exico\\
Ap. 70-543, 04510, Ciudad de M\'exico, M\'exico}

\begin{abstract}
We have studied the evolution of HuBi\,1-like planetary nebulae, considering several stages of mass injection. We have carried out numerical ionization+1D hydrodynamics+atomic/ionic rate models with our code {\sc Coral1d} to reproduce planetary nebulae that present multiple shells produced by different ejection events around the ionizing source. We are interested in comparing numerical simulations with H$\alpha$ and [\ion{N}{2}]$\lambda$6584 emission structures and the position-velocity diagrams observed in HuBi\,1. This object also has a phase where it has drastically decreased the injection of ionized photons ejected from the source. The result of these different stages of ejection is a nebula with intense [\ion{N}{2}] line emission in the inner part of the planetary nebula and an extended \ion{H}{2} recombination line emission around the central zone.
The model for HuBi\,1  shows the capability of our code to explain the hydrodynamical and photoionization evolution in ionization nebulae. This is our first step with a 1D code to study these two physical phenomena at the same time. 
\end{abstract}

\section{introduction}
Planetary nebulae (PNe) are ionized gas envelopes that surround evolved low-intermediate mass stars. This gas was originally part of the stellar envelope. The nebular ejection processes occur during the red giant (RG) and Asymptotic Giant Branch (AGB) epochs and produce nebulae of different types of morphology, density distribution, chemistry, and kinematics depending on the characteristics and conditions of the progenitor star. The physical characteristics of such processes are still debatable, and they depend strongly on the stellar properties, being very different if the star is single, is in a detached binary system, is in a close binary system, and/or if it possesses an accretion disk.

Many PNe show evidence of several ejections at different epochs during the AGB and post-AGB phases; thus, PNe can present several shells of different morphologies, sizes, and other physical characteristics such as density, temperature, ionization degree, and even different chemistry.

On several occasions, it has been found that PN emission might be variable on short time scales or the PN might be ionized by a variable star (e.g., \citealp{balick:21}; \citealp{otsuka:17}; \citealp{hadjuk:15}; \citealp{pena:22}). 

In the  HR diagram, single central stars are expected to evolve from the AGB phase towards higher effective temperatures at nearly constant luminosity (see e.g., evolutionary models by \citealp{miller:16} and others). The maximum temperature achieved could be a few hundred thousand K depending on the initial and/or final stellar mass. Afterward, the star enters the cooling track of white dwarfs, causing the evolved PN to slowly recombine while dispersing. However, in some cases, the central star experiences a late thermal pulse (LTP) or a very late thermal pulse (VLTP), in which case the evolution path reverses, the stellar atmosphere expands and cools, and the star makes a loop in the HR diagram, briefly returning to the AGB zone before continuing its evolution towards the PN phase a second time. Central stars that have experienced such an episode are called {\it born-again} (see e.g., \citealp{iben:95}; \citealp{herwig:99}; \citealp{miller:06}, \citealp{Montoro-Molina:22}  and references therein).
 It has been proposed that in returning to the PN stage, the star becomes a Wolf-Rayet, [WR], star and a H-poor  nebula is ejected. However, not all the known [WR] central stars are {\it born-again}. It is tempting to propose that {\it born-again} stars  are produced as a result of binary evolution. However, PNe are produced by low-intermediate mass stars and multiple lines of evidence point to binary systems as the birthplace of  most PNe,  then it should be expected that several of the eight so far known {\it born-again} stars belong to a binary system, although until now, only A\,30 has been proposed as a possible binary star \citep{jacoby:20}. As already said, so far only eight objects have been identified as {\it born-again} stars. In all these cases, it is crucial to analyze the evolution of the physical parameters of the nebula and its dynamical behavior, trying to understand the stellar evolution, which is not always directly observable due to the faintness of the central star.\\

HuBi\,1 (PN G012.2+04.9, PM\,1-188, IRAS\,17514-1555) is one of the PNe whose central star apparently had a VLTP and presented a recent {\it born-again} episode (see \citealp{guerrero:18}). \citet{hubi:90} discovered that the central star of HuBi\,1 showed an unusually late [WC\,10] stellar type, and later studies determined  that it possesses an effective temperature of about 35,000 K (\citealp{uwe:98}; \citealp{guerrero:18}). The star has been fading with time, losing about 10 mag in the optical in the last 40–50 years. The nebula has two shells, one external, photoionized, emitting H and He recombination lines and weak heavy-element collisionally excited lines, and an inner shocked shell emitting intense collisionally excited lines of N$^+$, O$^+$ and S$^+$. 
 \citet{pollacco:94} were the first to mention that HuBi\,1 consists of two shells with the inner, very compact one (diameter smaller than 1.5 arcsec), much denser and brighter than the outer one. This structure was confirmed by \citet{pena:05}  and \citet{guerrero:18} who showed the existence of this inner nebula in HuBi 1 based on images of H$\alpha$ and [\ion{N}{2}]$\lambda$6584. They reported that the inner nebula has an inverse ionization structure typical of shocked gas. The inner nebula is responsible for the intense emission of [\ion{N}{2}] lines while the outer nebula emits mainly H recombination lines. \citet{rechy:20} and \citet{pena:21}  analyzed the nebular kinematics which shows the presence of an accelerated structure (with velocity larger than 100 km s$^{-1}$) in the inner region of the nebula with characteristics of shocked gas.
 
\begin{figure}
    \centering
    \includegraphics[width=0.9 \columnwidth]{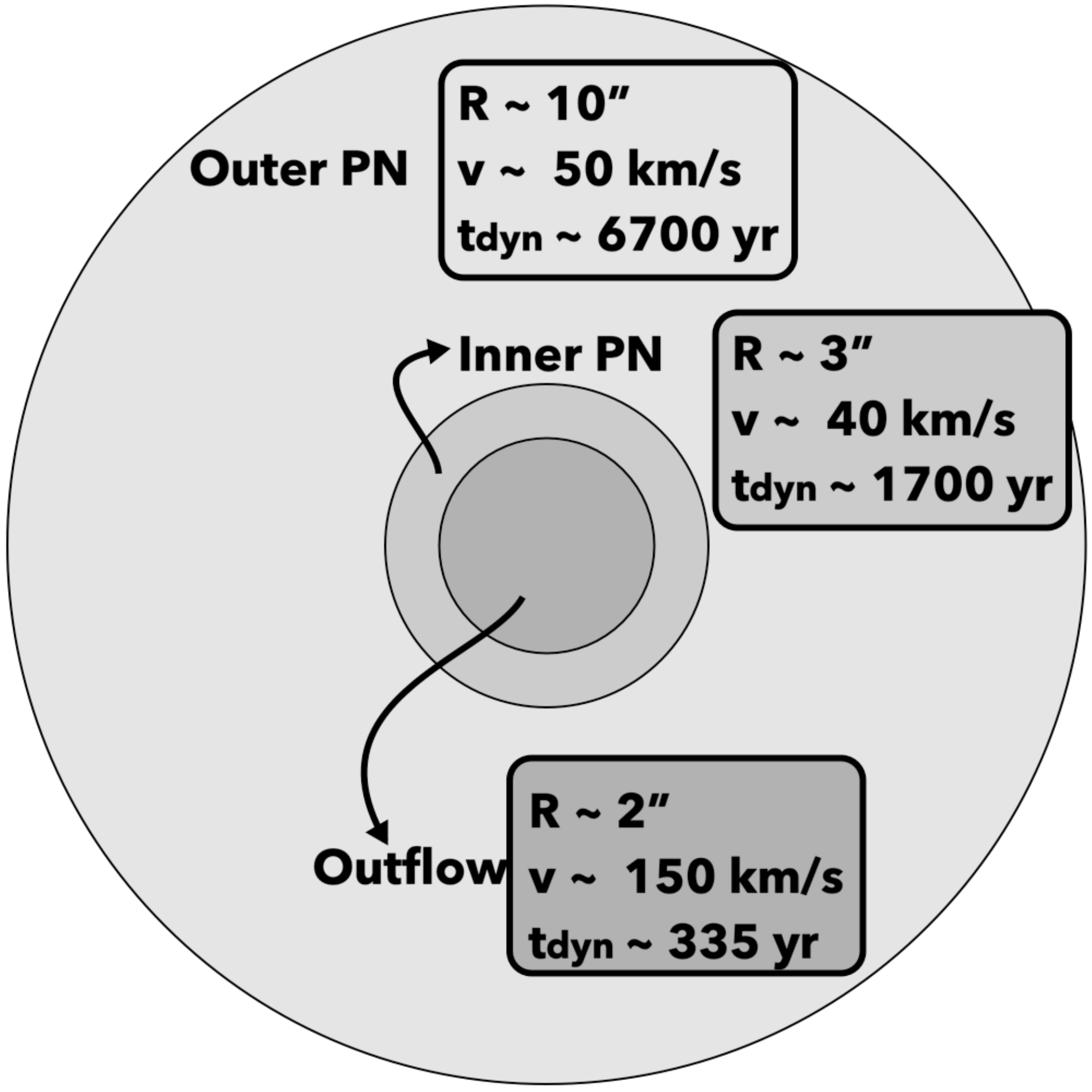}
        \includegraphics[width=0.9\columnwidth]{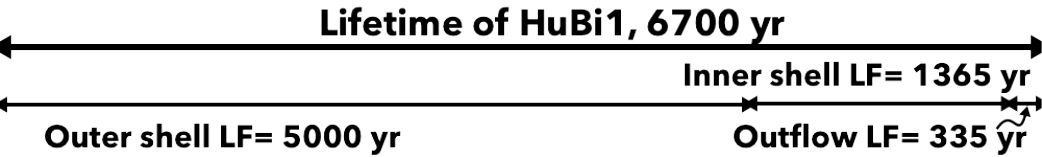}
   \caption{The outer and inner shells of HuBi\,1 and the inner outflow are schematically represented. Sizes of the shells (R), expansion velocities (v) and kinematic ages (t$_{dyn}$)  are taken from \citet{pena:21} (see \S 2).}
    \label{fig:scheme}
\end{figure}

The nebula is schematically represented in Fig. \ref{fig:scheme} where the nebular radii, expansion velocities, and kinematic ages of the different components, as presented by \citet{pena:21}, have been included. Kinematic ages depend on the distance and radii adopted (different authors mention different sizes depending on the depth of their images). For HuBi\,1 the heliocentric distance is  6.88$\pm$2.19 kpc, as derived by \citet{frew:16}. Ages therefore have large error bars, however the values we are using in Fig.  \ref{fig:scheme} are in agreement with the values reported by \citet{rechy:20} and \citet{pena:21}. 

In this paper, we use hydrodynamic simulations to try to replicate the particularly intriguing characteristics of HuBi\,1. 
Numerical simulations have become essential tools in astrophysics. Hydrodynamical models have been computed since long ago with the aim of understanding the physical processes of ejection and evolution of planetary nebulae by considering the stellar and environmental characteristics (\citealp{balick:02}; \citealp{raga:00}; among others). One 3D radiation-hydrodynamic simulation for HuBi\,1 was generated by \citet{toala:21}, by adopting mass-loss and stellar wind terminal velocity values obtained from observations.  The model by these authors assumes that the inner shell of HuBi\,1 was formed as a result of a very energetic VLTP which ejected H-poor material with velocities of about 300 km s$^{-1}$, 200 yr ago. They concluded that the large variations in the stellar wind parameters (mass ejection and the terminal velocity) produce instabilities and these  instabilities can form structures that can be seen as clumps and filaments of ionized material between the shells of HuBi\,1. This ionized gas produced the [\ion{N}{2}]$\lambda$6584 emission.  However, \citet{toala:21} have not a photoionization code or the atomic/ionic rates, therefore they could not directly conclude about the formation of emission lines produced by shocks, for example, the emission of [\ion{N}{2}]$\lambda$6584. Moreover, the results they have obtained are only related to variations in velocity and mass injection rate during the VLTP phase. The temporal distance between the events is not entirely clear. This is because from the observations only the kinematic times are available and the formation of internal structures will be more dependent on the time between the events that allow for the interaction between the shells.

  
This paper is organized as follows: In \S1 an introduction describing previous work on HuBi\,1 is included. In \S2 the numerical simulations for building the hydrodynamical models are introduced. In \S3, we present the numerical models. Our results regarding the density structure and gas properties (velocity, pressure, and temperature structures) can be found in \S4.  In \S5 we present our conclusions.

 \section{Numerical simulation}
We developed a set of numerical simulations using a hydrodynamical spherically symmetric code named {\sc Coral1D}, based on the code presented in \citet{mellema:98} and \citet{raga:00}, where the photoionization and non-equilibrium radiative cooling functions are considered. The code solves the mass, momentum, and energy conservation equations.

\begin{equation}
\frac{\partial \mathbf{U}}{\partial t}+\frac{\partial \mathbf{F}}{\partial
R}=\mathbf{S},
\end{equation}
where $\mathbf{U}$ is the vector containing the conservative variables which consider several ions of H, He, C, O, Ne, S and N which are calculated as advection equations. The vector of conserved variables is formed by
\begin{equation}
\begin{split}
       \mathbf{U} = &[E, \rho u, \rho, n_{HI}, n_{HII}, n_{HeI} n_{HeII}, n_{CIII},
         n_{CIV}, n_{CV}, \\ &n_{OII}, n_{OIII}, n_{OIV}, n_{OV},n_{OVI}, n_{NeII},
       n_{NeIII},n_{NeIV},\\ &n_{NeV},n_{NeVI},n_{SIII}, n_{SIV}, n_{SV}, n_{SVI},
   n_{NIII}, n_{NIV}, \\ & n_{NV}, n_{NVI}],
\end{split}
\label{eq:u}
\end{equation}
where $\rho$ is the density, $u$ is the velocity in the radial direction $R$, $E=\frac{1}{2}\rho u^2+\frac{n k T}{\gamma -1}$, is the internal energy, $n$ is the density, $T$ is the temperature giving by state equation $P=(n+n_e)k T$ ($n_e$ is the electron density), and $\gamma$ is the ratio of heat capacities. The vectors are filled with the densities of the different ions, $n_{ion}$.  $\mathbf{F}$ is the fluxes vector and is given by 
 \begin{equation}
    \mathbf{F}=[\rho u,\rho u^2+P,  u (E+P), n_{HI} u, ...,n_{NVI} u ],
    \notag
\end{equation}
and  $\mathbf{S}$ is the sources vector. Therein, we calculate the heating and cooling rates and the ionization and recombination coefficients  due to  the photoionization and the emission of radiation. The atomic parameters used in {\sc Coral1D} are described in \citet{raga:97}.

Our main aim in this work is to model PNe that have had more than one episode of matter ejection. In particular, we are interested in studying planetary nebulae that have presented ejection episodes with important differences in the momentum flux injected between episodes.  As an example, in this paper we are going to focus on HuBi\,1, a well known PN, recently analyzed by \citet{guerrero:18} and \citet{pena:21}, among others,  with at least three distinct episodes of matter ejection, as described below: 
\begin{enumerate}
    \item [a)] An external nebula, with a present radius of $\sim$10 arcsec as derived from the position-velocity diagram (PVD) presented in Fig. 6 by \citet{pena:21} and Fig. 1 by \citet{guerrero:18}. It shows an expansion velocity of 50 km s$^{-1}$, and it was ejected about 6700 yr ago. The gas is photoionized, and in the process of recombination, due to the systematic weakening of the central star.
    \item [b)] An inner nebula with a radius of $\sim$3 arcsec (same references as in item (a)), and an expansion velocity of 40 km s$^{-1}$, showing  intense [\ion{N}{2}]$\lambda$6584 emission that is evidence of being shocked. It was ejected about 1700 yr ago.
    \item [c)] An outflow or jet immersed in the central part of the inner nebula with a size of about 2 arcsec and a velocity of more than 150 km s$^{-1}$;  \citet{rechy:20} claims a maximum velocity of 250 km s$^{-1}$ for this internal outflow. It seems to have been ejected about 335 yr ago.
\end{enumerate}

\subsection{Emissivity and  column density maps}

\begin{figure}
    \centering
    \includegraphics[width=0.95 \columnwidth]{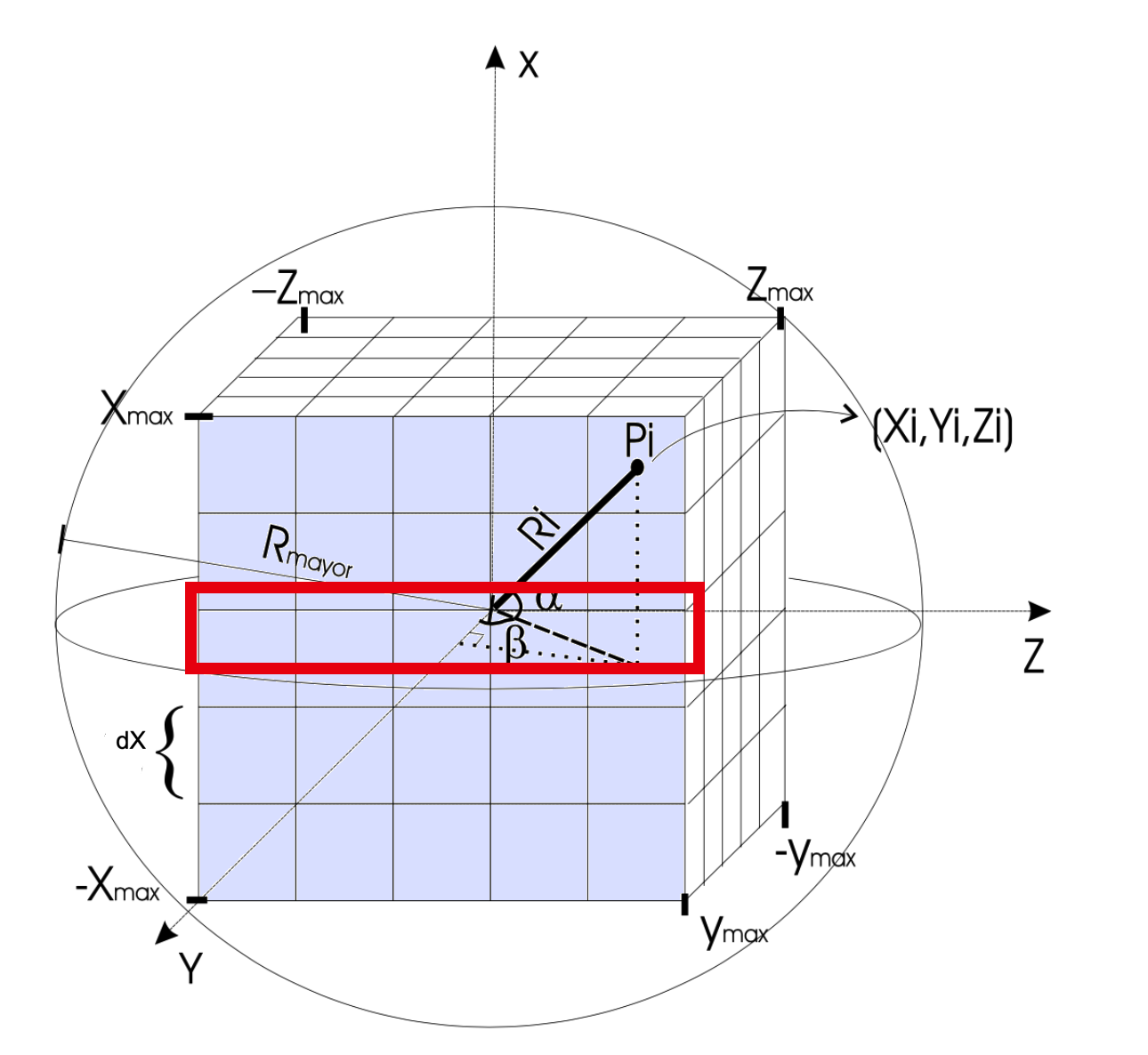}
    \caption{Cubic grid used for the model. The red rectangle represents a 2 arcsec slit through the center.}
    \label{fig:grid}
\end{figure}

 To create column density and emission maps, we have inserted the numerical results into a cubic grid (see Figure \ref{fig:grid}). The cells of this grid have a position P$_i$ with coordinates (X$_i$, Y$_i$, Z$_i$), and with a distance to the center of the simulation given by R$^2_i$=X$^2_i$+Y$^2_i$+Z$^2_i$.
(X$_{\rm max}$, Y$_{\rm max}$,Z$_{\rm max}$) is the maximum size of our simulation box.  To fill the grid, we have interpolated the vector $\mathbf{U}$ from the equation (\ref{eq:u}) using the R$_i$ of the cells.

We have derived the emissivity coefficients of H$\alpha$ and [\ion{N}{2}]$\lambda$6584 using the physical properties of each of the cells in the 3D grid (i.e. the temperature and the ionic and electron densities). 

Projected maps for H$\alpha$ and [\ion{N}{2}]$\lambda$6584 were obtained by integrating the emissivity coefficients at different lines of sight. We have also constructed the column density maps for H$^+$ and  N$^+$ to help us explain the  H$\alpha$ and [\ion{N}{2}]$\lambda$6584 maps.  

 \subsection{The H$\alpha$ and [\ion{N}{2}]$\lambda$6584 emissivities} \label{sec:emis}
 The emission lines predicted by the models are calculated using the temperature and density of each of the cells. For the H$\alpha$ emission, we have considered the parametric given by, 
%
\begin{equation}
    j_{H\alpha}=\frac{1}{4 \pi} \left [\chi_{ \rm HII}^2 n_H^2(\chi_{ \rm HII}\; \epsilon_r({\rm H}\alpha)+(1-\chi_{ \rm HII} \, \epsilon_c ({\rm H}\alpha_c))\right].
\end{equation}
where, $\epsilon_r(\rm{H}\alpha)$ and $\epsilon_c (\rm{H}\alpha)$ are the parametric recombination and collisional coefficients fitted by  \citet{raga:15} and  $\chi_{ \rm HII}$ is the ionization fraction.\\

For the [\ion{N}{2}]$\lambda$6584 emission line, we are using the coefficients obtained by solving the statistical equilibrium equation for an atom/ion of 4 levels. The collisional strengths coefficients and Einstein coefficients 
considered are given by \citet{aller:84}, and are dependent on the temperature and the electron density, $\Omega(n_e, T)$. Therefore, the [\ion{N}{2}]$\lambda$6584 emission coefficient is given by, 
\begin{equation}
    j_{[N\,II] \lambda 6584}=\frac{1}{4 \pi} f(n_e, T) n_e n_{NII}
\end{equation}
where $n_{N\,II}$ is the numerical density of  \ion{N}{2} in each cell.

\subsection{Position-velocity diagrams}
To compare our numerical simulation with the observational data, we have built  position-velocity  diagrams (PVD), integrating the emission of the maps for different radial velocities. They were calculated by convolving the integrated H$\alpha$ and [\ion{N}{2}]$\lambda$6584 emission maps with the Doppler profile of the emission line. We have considered 210 velocity channels from $-$150 to 150~km s$^{-1}$, which is the maximum velocity used in the models. 
To compare the PVDs shown in the observations with our diagrams, we have included the effect of a 2 arcsec wide slit through the center of the simulation box.
Such a PVD was built for the last model and is presented in \S 4.

\section{Numerical Models}
All our models consider that there was a mass injection during the AGB phase of the progenitor star of the planetary nebula. The duration of the AGB phase is sufficiently long to fill the entire simulation box with gas that was injected at a velocity of 10 km s$^{-1}$, a mass injection rate of 1$\times$10$^{-5}$ M$_\odot$/yr, and a temperature of 200 K, typical of observed AGB winds \citep{kwok:78}. The ejected gas has solar abundance \footnote{\citet{kwitter:22} have compiled the abundances of a large amount of PNe presented in several disk PN surveys and concluded that mean values of O, Ne and Ar abundances are close to solar, while He, C, N abundances are enhanced only by a few tenths of a dex, thus we have adopted solar abundances for the ejected gas in our models.}.  The simulation box is 0.325 pc per side, in a one-dimensional grid of 4000 solution cells,  $\sim$ 16.72 au/pix. Using this simulation, we have constructed 2D projection maps of 500 cells per side, with a resolution of 131 au/pix.

The planetary nebula ejection has been divided into 3 stages: 

1) First, a stage where the star has a luminosity of 10$^{3.8}$ L$_\odot$  and an effective temperature of 35,000 K \citep{guerrero:18}. In this stage, a stellar wind is produced with a mass injection rate of 2$\times$10$^{-6}$ M$_\odot$/yr and a velocity of 360 km s$^{-1}$ \citep{kwok:78}. This stage is injected at a time t = 0 yr. 

2) In the second stage, temperature and luminosity do not change with the first stage, as expected, but the wind that will form the inner nebula has a greater mass injection rate:  1$\times$10$^{ -5}$ M$_\odot$/yr
 and a velocity of 50 km s$^{-1}$.  This stage is injected at a time $t=\Delta t_{in}$ (see Table \ref{tab:models}).

3) Finally, there is a last stage with a significant decrease in the injection of ionizing stellar photons, so we consider that the number of photons injected is 4 orders of magnitude lower than for the previous stages. This is consistent with the fact that the central star of HuBi\,1 has declined by more than 10 mag in the visual, in the last 40$-$50 years. The wind injected at this stage has a velocity of 150 km s$^{-1}$ and a mass injection rate of 2$\times$10$^{-6}$ M$_\odot$/yr (like the first stage). This stage is injected at a time $t=\Delta t_{in} + \Delta t_{out}$ yr (see Table \ref{tab:models}).

Table \ref{tab:models} shows the model properties, including the previous wind injected by the star in the phase of AGB, and the three different outflows that are considered in this work. 

It is important to note that this study is only focused on quantifying the importance of each injection episode.  The velocities and mass injection rates used in each of the episodes were selected to reproduce the observed kinematics in the three regions of the nebula HuBi\,1, namely, the outer, inner, and jet nebula, as well as the currently observed velocity. To achieve this, we performed a series of simulations that allowed us to adjust these values within  intervals proposed by a standard planetary nebula range from a hundred to a thousand km~s$^{-1}$ (e.g., \citealp{kwok:78}; \citealp{perinotto:04}; etc.). For the outer PN  and for the inner PN nebula, we have explored velocity and mass loss rate values for the VLTP stage, such as those presented in \citet{GG:2020} and \citet{toala:21} since tens to a hundred km~s$^{-1}$. It is clear that there are other possible combinations of velocities  and mass loss rates, for each episode, that can give similar results to those presented in this work. However, the physical parameters are within the expected intervals for each of the stages and the dynamic results we obtain with them show a good agreement with the observed values for this object. A more comprehensive study of velocities and mass loss rates for each stage can be conducted in future works, provided that the observational constraints we have for this object are considered, but this is outside the goals of this paper.

\begin{table}[h]
\begin{center}
\caption{Physical initial conditions of the numerical simulations} 
\label{tab:models}
{
 \begin{tabular}{l c c cl}
\hline \hline
\multicolumn{1}{c}{Model} &
\multicolumn{1}{c}{$\Delta$ t$_{\rm {in}}$} &
\multicolumn{1}{c}{$\Delta$ t$_{\rm {out}}$}\\

\multicolumn{1}{c}{} &
\multicolumn{1}{c}{[yr]} &
\multicolumn{1}{c}{[yr]} \\
\\
\hline
M0$\ddag$ & 6500 & $--$ \\ 
M1$\dag$ & 5000 & 1200 \\ 
M2 & 5000 & 900 \\ 
M3 & 5000 & 1000 \\ 
M4 & 5000 & 1200 \\ 
\hline
\hline

\end{tabular}
}
\end{center}
\footnotesize{$\ddag$ the model considers a single ejection event that remains constant throughout the evolution time, and $\dag$ the model considers two ejection events, with different ejection physical properties.}
\end{table}

As we have said before, from observations it is found that PNe can have   many injection stages, however, initially, we have explored the simplest models, models M0 and M1, in which not all the observed injection stages mentioned above are considered.

Our first simplest developed model, M0, considers a constant mass injection rate of 2$\times$10$^{-6}$ M$_\odot$/yr,  an ejected gas velocity of 360 km s$^{-1}$, and constant effective temperature and luminosity during all lifetime of 6500 yr. 

In model M1 we study the effect of the second stage of the material ejection, considering an increase in the mass injection rate and a decrease in the material injection velocity from the central source, as mentioned above.

Finally, to study the effect of the third ejection stage  on the evolution of the planetary nebula, we have included a variation in velocity, mass injection rate and the number of photons injected by the central source (in the form of luminosity and effective temperature). For this stage, we consider that the star emits no ionizing photons and therefore, during the last 300 yr of evolution of this object, we have decreased the luminosity and effective temperature of the ionizing source by 10 orders of magnitude. We present the results of 3 models, M2, M3 and M4, for which we have only changed the duration of the second stage (900, 1000, and 1200 yr respectively). It is true that the observational articles propose a (kinematic) age of 1700 yr, which corresponds to the maximum age of the inner nebula. With models M2, M3 and M4, we explore the effect of this last ejection stage on the emission maps obtained from our simulations and compare them with the observations.

\section{Results}

\subsection{Model M0}
\begin{figure}
    \centering
    \includegraphics[height=0.9\columnwidth]{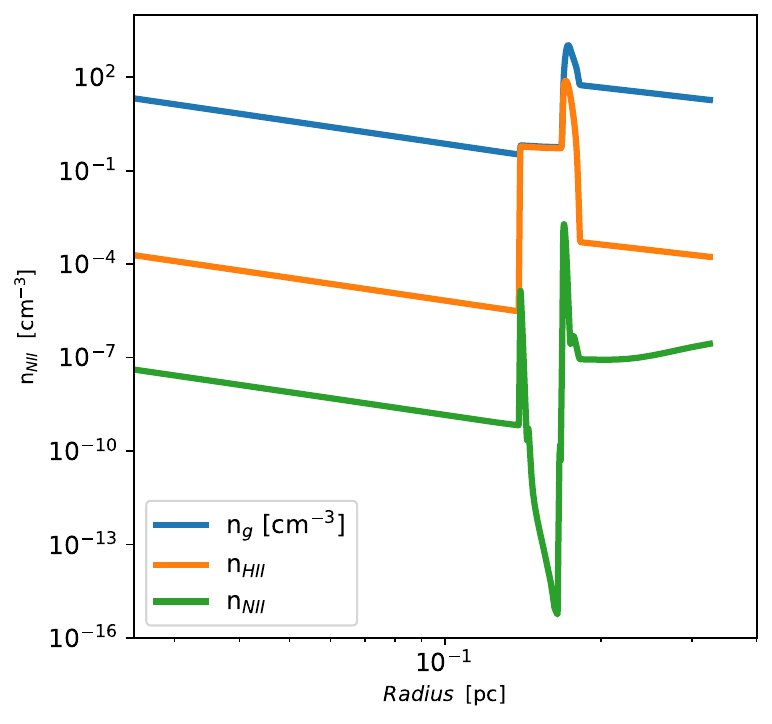}
    \caption{ Gas (blue), \ion{H}{2} (orange) and \ion{N}{2} (green) densities for model M0,  as a function of radial distance, for an evolutionary time of 6500  yr, respectively.}
    \label{fig:M0:dens}
\end{figure}

In Figure \ref{fig:M0:dens}  we present the numerical densities of the gas (blue),   \ion{H}{2} (orange) and \ion{N}{2} (green), as a function of the PN radius,  for model M0 at 6500 yr. 

In this figure we can see the frontal or main shock that interacts with the AGB wind that moves at a speed of 10 km s$^{-1}$ and has a low temperature of 200 K. The gas, shocked by the frontal shock, has a density that is at least two orders of magnitude larger than the AGB wind it encounters. We can also appreciate a more internal structure, at $\sim 0.17$ pc, which is the gas ejected by the star that is in a reverse shock that is balancing pressures in the shell. This same structure is shown in the density of \ion{H}{2}. However, this density structure is not seen in the nitrogen ion, \ion{N}{2}. In this latter case, although the increase in compression in the rear of both shocks is observed, there is also a drastic decrease in the density of \ion{N}{2} between the front and reverse shocks. This
``hole" in the \ion{N}{2} density is due to the high temperature that prevails in much of the region between the shocks; in this zone, the nitrogen is more highly ionized.

\begin{figure}
    \centering
    \includegraphics[height=0.7\textheight]{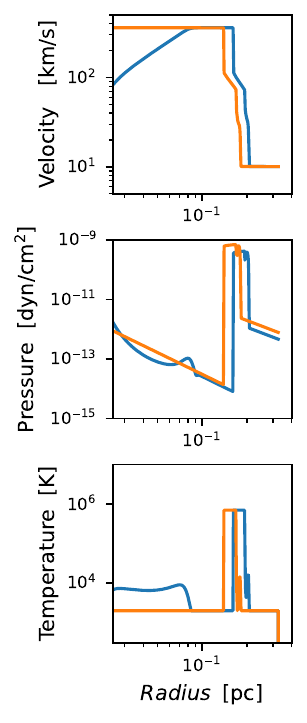}
    \caption{Gas velocity, gas pressure and temperature  as a function of radii  (upper, middle and bottom panel, respectively), for model M0 and M1 (orange and blue line respectively).} 
    \label{fig:M0:prop}
\end{figure}
Figure \ref{fig:M0:prop} shows the gas properties (velocity, pressure and temperature) as a function of the radial coordinate. The velocity distribution presents 3 zones (from model M0): a) the injection zone (off left side of the simulation box) at 360 km s$^{-1}$, b) the shocked gas injected in the first stage, with a velocity of about 70$-$90 km s$^{-1}$ and c) the shocked interstellar medium  with a velocity of 30$-$40 km s$^{-1}$. 
The gas shocked by the main shock and the reverse shock presents a significant increase in pressure (see middle panel of Figure \ref{fig:M0:prop}) that reaches up to almost 10$^{-9}$ dyn cm$^{-2}$. Same as with the pressure, a strong increase in temperature can be observed due to the shock heating. The temperature reaches a maximum value of near $10^6$ K, which corresponds to the post-shock temperature of a shock propagating with a velocity of a few hundred km s$^{-1}$. This region will emit mainly in X-rays. 

Furthermore, in the frontal part of the nebula, the temperature of the gas is  $\sim$ $10^4$ K, which corresponds to a photoionized region, which has mainly optical emission. 

\subsection{Model M1}
We now analyze model M1 which has 2 different mass injection events. The first one is identical to that  in model M0. The second starts at 
t = 5000 yr with an injection velocity of 50 km s$^{-1}$ and mass injection of $10^{-5}$ 
M$_\odot$/yr. We have kept
 the stellar luminosity and effective temperature constant for the entire time of the simulation.
  Figure \ref{fig:M0:prop} shows the time evolution of gas velocity, pressure and temperature of model M1 (in the blue line). In the upper panel, we can identify the same 3 velocity zones as in model M0, but in the injection zone, there is a flow with a lower velocity, producing a gradient of velocities towards the innermost region of the nebula.

Although the second stage injects material with a smaller velocity, this in turn has a greater injection of mass. In the middle panel of Figure \ref{fig:M0:prop} it can be seen that this combination produces structures in gas pressure that move inwards.  In the lowest panel of the figure, which shows the temperature, it can be seen how these structures correspond to shock waves, moving at hundreds of km s$^{-1}$, that heat the gas, producing a region with optical emission in the inner part of the nebula. 
 
\subsection{Models M2, M3, M4}

To study the effect of the last phase where the mass-loss rate has recovered a value of 2$\times$10$^{-6}$ M$_\odot$/yr and a velocity of 360 km s$^{-1}$   but the star has faded and does not produce any ionizing photons, we have run models M2, M3 and M4, where the second injection stage lasts
 900,  1000 and 1200 yr respectively, following the two previous stages. Because in this last stage no ionizing photons are produced, the nebula is recombining and its ionization structure changes completely.
 
\begin{figure}
    \centering
    \includegraphics[height=0.7\textheight]{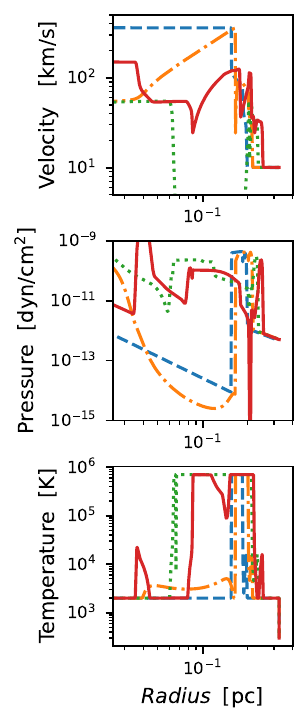}
    \caption{Gas velocity, gas pressure and temperature as a function of radius for model M2 (upper, middle and bottom panel, respectively). The blue (dashed), orange (dash-dotted), green (dotted) and red (solid) lines correspond to the evolutionary time of 5000, 5500, 6000 and 6500 yr, respectively.}
    \label{fig:M2:prop}
\end{figure}

For model M2, we allow the second stage to last only 900 yr. In Figure \ref{fig:M2:prop} we can see the velocity, pressure and temperature structures of the nebula at 5000, 5500, 6000, and 6500 yr. As expected, we find that the pressure structure at 6000 yr (green-dotted line) has considerable differences regarding that of model M1 (which does not consider the third stage), in particular, the pressure structure shows a maximum in the front part, corresponding to the outer part of the nebula where it is expected to have optical emission. It also presents another local maximum between 0.07 and 0.1 pc. This pressure gradient is reflected in a large drop in the velocity of the gas, due to the formation of an internal or reverse shock wave ( which can be seen as a jump in density due to the swept injected gas) that is thermalizing the gas injected by the star.  In fact, it is possible to observe the shocked region in the gas temperature plot, where it is observed that there is a temperature of $\sim$ 6$\times$10$^5$ K from 0.07 pc to 0.2 pc, which corresponds to a region crashed at a few hundred km s$^{-1}$. However, 500 yr later, at 6500 yr of evolutionary time (red line in Figure \ref{fig:M2:prop}), it appears that the shocked region has started to cool, yet other shock waves are being generated from the front of the nebula structure.  For this epoch, we can also see a structure with a temperature of $\sim$ 10$^4$ K, at 0.04 pc of the injection region, which corresponds to the size of the shocked [\ion{N}{2}] region that occurs in the inner zone of HuBi\,1 (2 arcsec).

\begin{figure}
    \centering
    \includegraphics[height=0.7\textheight]{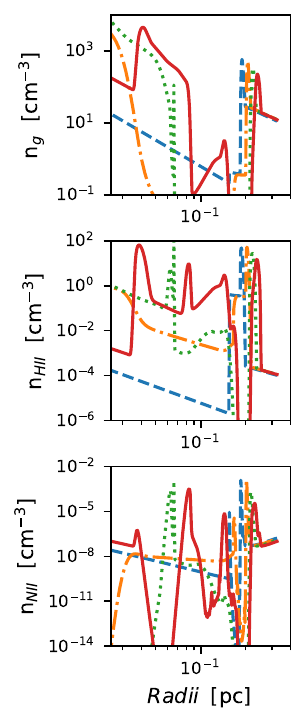}
    \caption{{Gas, \ion{H}{2} and \ion{N}{2} density as a function of radius for model M2 (upper, middle and bottom panel, respectively). Same  as in Fig. \ref{fig:M2:prop}.}}
    \label{fig:M2:density}
\end{figure}

Figure \ref{fig:M2:density} shows the density structure of gas, hydrogen and ionized nitrogen for model M2, at different evolutionary times of the planetary nebula. Clearly,  4 local maxima in the density of hydrogen and nitrogen can be seen at 0.04, 0.08, 0.12 and 0.21 pc, however, the only regions that can have a significant optical emission are the innermost and outermost regions where the temperature is $\sim 10^4$ K.

\begin{figure}
    \centering
    \includegraphics[width=1.0\columnwidth]{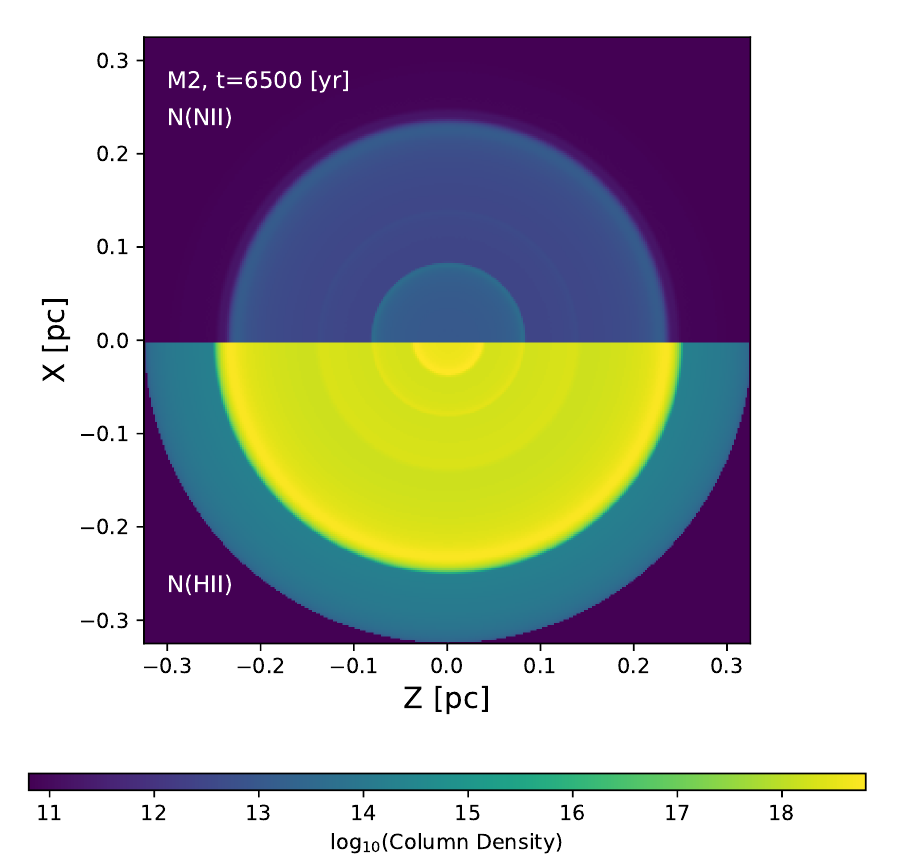}
    \caption{\ion{H}{2} and \ion{N}{2} column density, in the bottom and upper panel, respectively, of model M2, for the evolutionary time of 6500 yr.}
    \label{fig:M2_ColDens}
\end{figure}

\begin{figure}
    \centering
    \includegraphics[width=1.0\columnwidth]{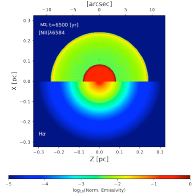}
    \caption{H$\alpha$ and [\ion{N}{2}]$\lambda$6584 emission, in the bottom and upper panel, respectively, of model M2, for the evolutionary time of 6500 yr.}
    \label{fig:M2_Emis}
\end{figure}

\begin{figure}
    \centering
    \includegraphics[width=1.0\columnwidth]{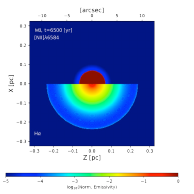}
    \caption{H$\alpha$ and [\ion{N}{2}]$\lambda$6584 emissivity, in the bottom and upper panel, respectively, of model M3, for the evolutionary time of 6500 yr.}
    \label{fig:M3_Emis}
\end{figure}

\begin{figure}
    \centering
    \includegraphics[width=1.0\columnwidth]{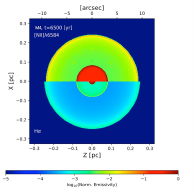}
    \caption{H$\alpha$ and [\ion{N}{2}]$\lambda$6584 emission, in the bottom and upper panel, respectively, of model M4, at a time of 6500 yr.}
    \label{fig:M5_Emis}
\end{figure}

To study this region in more detail, we have plotted \ion{H}{2} and \ion{N}{2} column densities, shown in the lower and upper panel in Figure \ref{fig:M2_ColDens} respectively.  Notice that our Figures~\ref{fig:M2_ColDens}, \ref{fig:M2_Emis}, \ref{fig:M3_Emis} {and \ref{fig:M5_Emis}} were reconstructed through the utilization of a one-dimensional simulation, utilizing spherical symmetry, to schematically depict the shells that are formed by each of the ions. As it can be seen, the column density of \ion{H}{2} reaches a maximum radius of 0.21 pc, and shows ``rings" at 0.04, 0.08, 0.12 pc and the outside at 0.21 pc, corresponding to the maximum values shown in the density plots. On the other hand, the \ion{N}{2} density shows only 2 significant rings, a very inner one at about 0.05 pc and an outer one at the external radius of the nebula. 

To compare the numerical result with the observations (shown in Fig. 1 of \citealp{rechy:20}), we have made maps of the line emission of  [\ion{N}{2}]$\lambda$6584 and H$\alpha$ in the upper and lower panel of Figure \ref{fig:M2_Emis}. We present the base 10 logarithms of normalized emission, where it is observed that the emission of [\ion{N}{2}]$\lambda$6584 from 0.07 pc to the outer radius of the planetary nebula is smaller than in the inner part, where it is up to 100 times larger. On the other hand, the H$\alpha$ map shows a drop in emission as a function of radial distance.

In the case of HuBi\,1 it is observed that the emission of H$\alpha$ is more extended than that of [\ion{N}{2}]$\lambda$6584, and this is not the case for the result of the model M2. For this reason, we have decided to explore, in model M3, the effect of starting the third stage a little later, for example at 6000 yr of evolutionary time, that is, the second stage lasts 1000 yr. 

A map of the [\ion{N}{2}]$\lambda$6584 and H$\alpha$ emissions of this model is shown in Figure \ref{fig:M3_Emis}, where the [\ion{N}{2}]$\lambda$6584  emission is concentrated in a compact $\sim $ 2$-$3$^{"}$ (0.05 $-$0.08 pc) region around the central source, while the emission of H$\alpha$, unlike the model M2, is more extended without reaching 10$^{"}$ (0.26 pc) like of the outer nebula of Hubi\,1.

To obtain an emissivity structure more similar to HuBi\,1, we have again increased the starting time of appearance of the third phase, and finally, we have found that model M4 where the second phase lasts more than 1200 yr, presents an extended  emission of H$\alpha$ in a region of $\sim$ 10$^{"}$ (0.26 pc, see Figure \ref{fig:M5_Emis}).  Such emission is relatively homogeneous (like the outer nebula of HuBi\,1), and there is an inner nebula of $\sim$ 3$^{"}$ (0.08 pc) where the emission from [\ion{N}{2}]$\lambda$6584 is much more intense (like the inner nebula of HuBi\,1, as well).

\begin{figure}
    \centering
    \includegraphics[height=0.75\columnwidth]{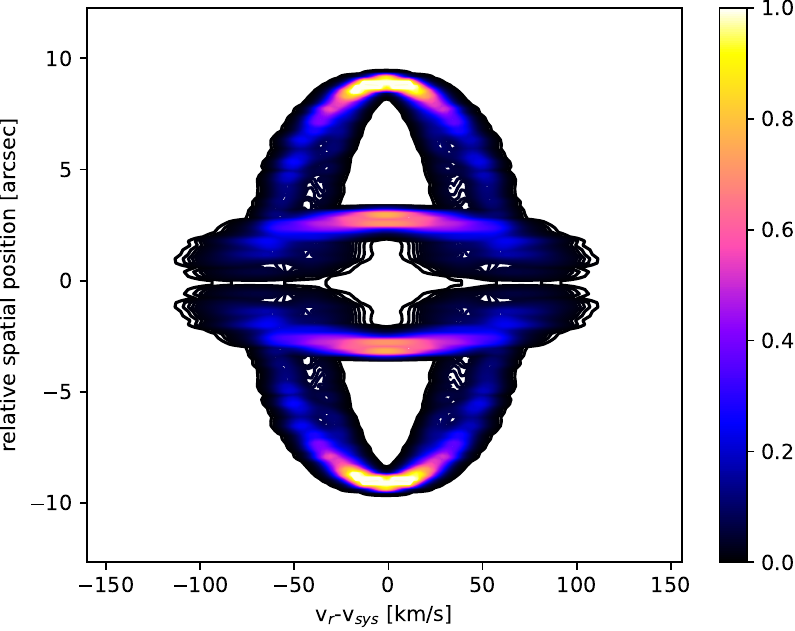}
    \caption{PV-diagram of the [\ion{N}{2}]$\lambda$6584 emission, of model M4, at 6500 yr.}
    \label{fig:pv}
\end{figure}

Finally, in Figure \ref{fig:pv}, we present the PV diagrams of the  model M4, for 6400 and 6500 yr, for the line [\ion{N}{2}]$\lambda$6584. As mentioned in section 2.3, the diagram was made by taking a slice of 2 arcsec (at 5.36 kpc which is the distance of HuBi\,1) of the image that is projected on the face of a cube. In this figure, we can observe the evolution of the 2 spherical shells of material injected by the planetary nebula. The innermost shell is the one that corresponds to the last injection phase that we have used in our models. Certainly, the internal part of our model evolves like a very fast shell, with speeds between $-$90 and 90 km s$^{-1}$  and sizes between 2 and 3 arcsec.

It is important to notice that the PV diagrams of our model cannot be directly compared with the PV diagrams observed by \citet{rechy:20} and \citet{{pena:21}} of HuBi\,1 because this object presents the last phase in which there is a no-spherical injection and the gas ejected might have a H-deficient chemical composition. In addition, we do not consider the possible reddening affecting mainly the receding zone of the PV diagrams. However, it is remarkable to observe how our models present the emission of the [\ion{N}{2}]$\lambda$6584 line in the inner part due to the shocks that interact with the material of the last phase.
%

\section{Conclusions}

We have presented a set of models for the evolution of HuBi\,1-like planetary nebulae, considering several stages of mass injection.  We have carried out 1D numerical simulations using a hydrodynamical spherically symmetric code named {\sc Coral1D}, where the  photoionization and non-equilibrium cooling and heating functions  are considered. In our models, the chemical abundances used for the plasma are solar. There was a previous work to model this object by \citet{toala:21}. However,  they have not included the non-equilibrium cooling and heating functions. Here we present a model that solves the hydrodynamics and the photoionization evolution of the gas,  capable of studying not only the hydrogen but other 25 species (ions and neutral). In this paper, we present the [\ion{N}{2}]$\lambda$6584 and H$\alpha$ lines to compare with the observations.

Our numerical models consider a previous AGB stage, where the medium has been structured by the supersonic gas blown during this stage. In this structure, there is a first stage of the planetary nebula in which wind-type material injected with a velocity of a few hundred km s$^{-1}$ and ionizing stellar photons emitted for approximately 5000 yr, forms  the outer nebula that appears in HuBi\,1. Then we added a new stage of a fast wind where the gas that is observed in the inner nebula will be injected for approximately 1000 yr. Finally, we injected the last stage where the central ``jet-like" object is injected with an injection velocity much higher than in stage 2, but we have turned off the ionizing photon source (as seen in HuBi\,1). 

In this work, we have shown that the drastic fading of the central star determines the observed structure of this type of object. This is because the decrease in photon flux translates into a reduction of temperature and pressure in the internal structures, creating holes in pressure and forming many shock waves that interact with the gas being ejected by the central object.

 As we mentioned above, we can not directly compare our PV synthetic diagrams obtained from the spherically symmetric models with the PV observed diagrams presented in \citet{{pena:21}}, but we can see that the gas from the latter ejection is shocked and is shown in the internal part, in the intense [\ion{N}{2}]$\lambda$6584 line, as predicted in \citet{{pena:21}}, therefore this result is in a good agreement with the observations. The main reason for not being able to compare the PVs is that the last stage of the gas evolution is complex and since our model is spherical, we lose the geometry. The observed object has a high speed ejecta in a cylindrical or jet shape. We can infer this from Figure 6 in \citet{{pena:21}}, where they show two slits at different positions (0$^{\circ}$ and 90$^{\circ}$) where we can compare their right upper panel (position at 90$^{\circ}$) with the high speed ejecta ($\sim$150 km s$^{-1}$) and their right lower panel (position at 0$^{\circ}$) with no high speed ejecta, this could give us a hint of the central object morphology as a ``belted outflow". Moreover, \citet{rechy:20} performed two ``pseudo-slits" (as they named them) and constructed the PV diagrams at two different angles 50$^{\circ}$ and 140$^{\circ}$, showing high velocity outflows ($\sim$200 km s$^{-1}$). If we join all these observed angles, we can picture a "belted outflow" shape. 

Finally, we present the model for HuBi\,1 as an example to show the capability of our model to explain the hydrodynamical and photoionization evolution in ionization nebulae. Although {\sc Coral1D} is a code presented by \citet{raga:08} with an excellent result, this is our first step with a 1D code to study these two physical phenomena at the same time. We are working on a 2D code to analyze different morphologies and different kinds of central ionization sources.


\begin{acknowledgments}
We acknowledge the support of the UNAM-PAPIIT grants IN110722, IN105020, IN111423, IG100422, and also the Miztli-UNAM supercomputer project LANCAD-UNAM-DGTIC-123 2022-1, and LANCAD-UNAM-DGTIC-128 2023-1. F.R.-E. and J.L.C. acknowledge scholarship from CONACyT, M\'exico. With great sadness we announce the passing away of our dear friend and colleague Dr. A. Raga during the writing of this paper. 
This work benefited greatly from his brilliant contribution in the field of hydrodynamics.
\end{acknowledgments}


\end{document}